\documentclass[conference]{IEEEtran}
\IEEEoverridecommandlockouts
\usepackage{cite}
\usepackage{amsmath,amssymb,amsfonts}
\usepackage{graphicx}
\usepackage{textcomp}
\usepackage{xcolor}
\def\BibTeX{{\rm B\kern-.05em{\sc i\kern-.025em b}\kern-.08em
    T\kern-.1667em\lower.7ex\hbox{E}\kern-.125emX}}
\usepackage{multirow} 

\usepackage{subcaption}
\usepackage{array}
\usepackage{stfloats}
\usepackage{multirow}
\usepackage{textcomp}
\usepackage{diagbox}
\usepackage{colortbl}
\usepackage{xcolor}

\begin{document}

\title{Performance Evaluation of Kubernetes Networking Approaches across Constraint Edge Environments}


\author{\IEEEauthorblockN{Georgios Koukis\IEEEauthorrefmark{3}\IEEEauthorrefmark{2},
Sotiris Skaperas\IEEEauthorrefmark{1}\IEEEauthorrefmark{2}, 
Ioanna Angeliki Kapetanidou\IEEEauthorrefmark{3}\IEEEauthorrefmark{2},
Lefteris Mamatas\IEEEauthorrefmark{1}\IEEEauthorrefmark{2},
Vassilis Tsaoussidis\IEEEauthorrefmark{3}\IEEEauthorrefmark{2}}
\IEEEauthorblockA{\IEEEauthorrefmark{2} Athena Research and Innovation Center, Greece}
\IEEEauthorblockA{\IEEEauthorrefmark{3} Department of Electrical and Computer Engineering, Democritus University of Thrace, Xanthi, Greece}
\IEEEauthorblockA{\IEEEauthorrefmark{1} Department of Applied Informatics, University of Macedonia, Thessaloniki Greece} 

Emails: \{gkoukis, ikapetan, vtsaousi\}@ee.duth.gr, \{sotskap, emamatas\}@uom.edu.gr 
}

\maketitle

\begin{abstract}

Kubernetes (K8s) serves as a mature orchestration system for the seamless deployment and management of containerized applications spanning across cloud and edge environments. Since high-performance connectivity and minimal resource utilization become critical factors as we approach the edge, evaluating the performance of K8s networking in this context is essential. This paper contributes to this effort, by conducting a qualitative and quantitative performance evaluation of diverse Container Network Interface (CNI) plugins within different K8s environments, incorporating lightweight implementations designed for the Edge. Our experimental assessment was conducted in two distinct (intra- and inter-host) scenarios, revealing interesting insights for both researchers and practitioners. For example, the deployment of plugins across lightweight distributions does not necessarily lead to resource utilization improvements, e.g., in terms of CPU/memory or throughput. 
\end{abstract}

\begin{IEEEkeywords}
Edge/Fog Computing, Kubernetes, CNI plugins
\end{IEEEkeywords}

\section{Introduction}

Edge and Fog computing architectures aim to bring computational resources closer to the user, enabling real-time processing and reducing latency. In this regard, Kubernetes (K8s) has emerged as a prominent solution 
for automated and scalable orchestration/management of containerized workloads and services. However, due to the resource constraints of Edge and Fog environments, a number of lightweight K8s implementations have been proposed, such as K0s, K3s, and MicroK8s,
that focus on reduced disk space, CPU and memory consumption \cite{Bohm}. In a similar vein, the ongoing project Cognitive Decentralised Edge Cloud Orchestration (CODECO, https://he-codeco.eu) investigates a K8s framework to enhance the flexibility of services across the edge-cloud continuum. CODECO pays particular attention on the identification of suitable toolkits for intelligent management of highly distributed environments with resource constraints.


Furthermore, the Edge 
holds immense significance in enhancing the networking communication of applications and devices that have stringent communication requirements, e.g., delay-sensitive IoT devices. In this context, plugins are responsible for the networking communication among the different K8s components, enabling their connectivity with overlay/underlay networks, e.g., through the Border Gateway Protocol (BGP), Virtual Extensible LAN (VxLAN), IP in IP, etc. A widely adopted plugin specification has been proposed by the Container Network Interface (CNI) project, which also comprises libraries for the 
configuration of networking interfaces. 

Although several works study the effect of CNI plugins e.g., \cite{Qi, Bohm} or lightweight K8s implementations e.g., \cite{Koziolek} on the resource utilization of clusters, the impact of networking plugins on resource-restricted environments, e.g., in the  Edge, is still under-investigated. We address this gap by conducting a qualitative and quantitative comparative analysis between various plugins across different K8s distributions, aiming to answer the following research questions:
\begin{itemize}
    \item RQ1: What is the impact of the interplay between the considered K8s distributions and CNI plugins in terms of resource utilization and network performance?
    \item RQ2: Can we associate the performance and functional characteristics of CNI plugins with particular edge application requirements? 
\end{itemize}

Along these lines, we assess $5$ representative CNI plugins across $4$ different distributions ($3$ lightweight and vanilla K8s) over two distinct testbeds. In particular, we measure the CPU and RAM utilization, and also the throughput gains of the compared CNI plugins across different resource-restricted environments and various connectivity conditions. These include idle, pod-to-pod, and pod-to-service conditions for both TCP and UDP communication protocols. Furthermore, we consider intra-host and inter-host communication scenarios, where hosts are located either on the same or on different machines. Our results reveal interesting insights regarding the two identified research questions.

The remainder of this paper is organized as follows. In Section II, we discuss recent related works that evaluate CNI plugins and lightweight K8s distributions. In Section III we approach the considered CNI plugins and distributions qualitatively. In Section IV we outline our experimentation setup and methodology, while in Section V we focus on the quantitative results of the experiments. Finally, Sections VI and VII detail the insights gained by our analysis along with our concluding remarks and future plans, respectively.

\section{Related work}

Recent literature mainly considers the networking impact of CNI plugins on the conventional vanilla K8s distribution. 
Authors in \cite{Qi} conduct a comprehensive comparison of various CNI plugins, focusing on functionality (e.g., packet forwarding and routing), performance and scalability aspects. They also investigate the impact of 
encapsulation and tunneling offload on the performance of selected plugins. 

Other relevant works, such as \cite{Kapočius2}, discuss the trade-offs between overlay and pure IP plugins, 
considering various Maximum Transmission Unit (MTU), Maximum Segment Sizes (MSS) and NIC offloading conditions. Paper \cite{Park} evaluates the throughput performance of the Flannel plugin, OVS-based and native-VLAN networks with different packet sizes, demonstrating Flannel's substantial computational overhead, especially as the MTU size decreases. A high-level resource-consumption-based assessment of various CNI plugins is carried out in the blog paper \cite{Ducastel3}. In addition, authors in \cite{Kapočius1, Zeng} assess the 
performance variations across virtual and bare metal environments, demonstrating the superior performance of the latter in terms of throughput and computational overhead.

Subsequently, several existing studies compare the performance of lightweight distributions, focusing on intra-distribution characteristics (e.g., execution time and resource consumption), which are crucial for their applicability in edge environments. In this context, \cite{Bohm}, \cite{Kjorveziroski} and \cite{Telenyk} investigate CPU, memory, and disk usage for MicroK8s, K3s and vanilla K8s distributions over various cluster life-cycle stages (e.g., startup, deployment and shutdown).
Surprisingly, the authors reveal that lightweight distributions do not always outperform the traditional K8s distribution. Paper \cite{Koziolek} extends the previous analysis with the performance comparison of K0s and MicroShift, while focusing on stress-inducing environments under generated workloads. Finally, the authors of \cite{Fathoni} quantify the performance of K3s and KubeEdge, in terms of CPU and RAM utilization, on a Raspberry Pi device.

Despite the broad interest, a research gap exists concerning the interplay between lightweight K8s distributions and different networking plugins. To the best of our knowledge, this is the first work to evaluate the performance of various CNI plugins across different K8s distributions. Moreover, we consider two distinct scenarios corresponding to nodes located on the same or on separate machines.

\section{Network plugins \& Kubernetes distributions}
In this section, we discuss the qualitative characteristics of the considered K8s distributions and CNI plugins.
Note that the selection of the specific plugins is determined by their compatibility with the discussed K8s distributions.

\subsection{Kubernetes distributions}
In particular, four well-known K8s implementations are considered, i.e., vanilla K8s, K3s, K0s and MicroK8s. 

Vanilla K8s (https://kubernetes.io) is the standard feature-rich and robust container orchestration platform. It is resource-intensive 
and suitable for large-scale production deployments (requiring a minimum of 2 CPUs and 2 GB RAM).

K3s (https://k3s.io) is a lightweight K8s distribution that focuses on resource-constrained environments, e.g., IoT and Edge devices. It requires minimal resources with half the memory of a vanilla K8s (even supports clusters with Raspberry Pi), combining components as a single binary ($<$100MB) with minimal to no OS dependencies. 
K3s replaces the default K8s storage with an SQLite-backed option (also supports Etcd3, MySQL, and Postgres), includes an ingress controller (Traefik), and operates with recommended system requirements of 2-core CPU and 1 GB RAM.

K0s (https://k0sproject.io) is a lightweight alternative of K8s packaged as a single binary ($<$160MB), allowing for easy installation across diverse environments. 
It prioritizes simplicity over features, creating separated control plane and deploying \textit{Konnectivity} agents 
to proxy traffic from the control plane (API server) to the worker nodes. K0s has minimum host OS dependencies, low system requirements (1 vCPU, 1 GB RAM) and minimal features compared to vanilla K8s and K3s, while it supports the same storing options as K3s.

MicroK8s (https://microk8s.io) is designed to run K8s on various Linux-based OSs, providing seamless installation through its snap package. It covers the needs for High Availability and CI/CD applications, including automatic updates and self-healing mechanisms due to snap refreshes. MicroK8s is suitable for both constrained environments, such as IoT and Edge scenarios, and high-demand production environments (with recommended requirements of 2-core CPU and 4 GB RAM). It uses dqlite storage  by default and provides additional features, such as load balancing, ingress controller (Traefik), dual-stack, GPU acceleration, observability and metric collection tools. 


\subsection{CNI plugins}
Additionally, we discuss the key features of the five selected CNI plugins, i.e., Flannel, Calico, Cilium, Kube-router and Kube-ovn.

Flannel (https://github.com/flannel-io/flannel) is a lightweight, Layer 2/3 overlay network fabric, which deploys a single binary agent (\textit{flanneld}) to assign an IP range of subnet addresses for each node in the cluster. It uses K8s API or etcd to maintain a mapping between allocated subnets and actual node IP addresses, while also supporting a variety of backend mechanisms, such as VxLAN (default L2), host-GW (L3), WireGuard, UDP, and other experimental options. However, it provides limited advanced features, e.g., security/encryption features and network policies. Flannel is the default CNI plugin for the K3s distribution.

Calico (https://www.tigera.io/project-calico/) 
is a feature-rich Layer-3 network solution. By default, it utilizes BGP mode, establishing a full BGP mesh among cluster nodes, while also supports alternative modes including IP overlay, VxLAN and Wireguard. Calico offers various kube-proxy modes such as iptables, ipvs, while it accesses etcd cluster directly or uses the K8s API datastore. Calico provides advanced network policy, administration, and security features, including Wireguard, eBPF (extended Berkeley Packet Filter), and XDP (eXpress Data Path). These features enable network policy enforcement (e.g. integrating with service meshes like Istio (https://istio.io), service discovery, load balancing and dual-stack networking. It functions as the default plugin in the MicroK8s distribution.

Cilium (https://cilium.io/)
is a Layer-3 networking, security and observability solution based on eBPF. It provides advanced programmable features, including network policy enforcement, transparent network security (with IPSec tunnels, TLS, GCM encryption), traffic monitoring and visibility (e.g., through the Hubble platform).
Cilium uses bpfilter routing instead of iptables, shifting filtering tasks to the kernel to improve performance, load balancing and processing of traffic speed.

Kube-router (https://www.kube-router.io/) 
is implemented at Layer 3, leveraging the standard Linux networking stack. It offers both overlay (IP-in-IP) and underlay (BGP) options, with the BGP protocol employed as the default in-cluster networking mechanism, utilizing the ``simple" bridge plugin and storing networking states in the K8s API. Deployed as a single binary solution, Kube-router supports network policies and utilizes LVS/IPVS for service proxy. It offers advanced features, such as Direct Server Return (DSR) and Layer-3 load balancing. 
It serves as the default K0s distribution plugin.

Kube-ovn (https://www.kube-ovn.io/)
is an SDN-based solution integrating the OVN-based network virtualization with K8s, providing network policies, subnet isolation, dual-stack support, and features like dynamic QoS, embedded Layer-2 load balancing and hardware offloading. It employs overlay networking by default, utilizing the Geneve kernel module to encapsulate cross-host traffic. However, Kube-ovn also supports underlay and Vlan networking modes for direct connectivity with the physical network and the integration with other plugins, e.g., Cilium.



To summarize the qualitative characteristics of the above plugins, Flannel provides a lightweight overlay solution, focusing on easy deployment and compatibility. However, it is not appropriate in deployments with scalability and security requirements, as well as with a demand for advanced network policies. Calico and Cilium are feature-rich and production-oriented solutions, with the former providing high performance and low latency, and the latter leveraging eBPF for programmable security and visibility features. Kube-router operates similarly to Calico, without the advanced features of the latter, while Kube-ovn provides ``rich" SDN capabilities.

Among the lightweight distributions, K3s and K0s reduce the OS footprint by minimizing or replacing certain resource-intensive K8s features. K0s focuses on an isolated control plane with limited features, while MicroK8s can be employed in both resource-constrained and production environments.

\section{Experimentation Setup and Methodology} \label{Section IV}
This section provides an overview of our experimentation approach, including the description of our setup and specifications of the considered experimentation environments. 

The installation of the K8s distributions and the respective CNI plugins, i.e., based on their most recent stable versions, 
is carried out in a cluster comprising one master and two worker nodes (VMs). Each node consists of 40GB storage, 4GB memory and 4 core CPU, running Ubuntu 22.04.2 LTS with kernel version 5.15.0-71-generic. 
The K8s cluster is being built based on kubeadm (v1.28.2) and Ansible, utilizing the ClusterSlice platform \cite{clusterslice}. 
After each scenario, the cluster is deleted and then re-deployed to remove any remnants of previous experiments. 

For the quantitative performance evaluation of CNI plugins, we employ the Kubernetes Network Benchmark (knb) tool (https://github.com/InfraBuilder/k8s-bench-suite), where the two worker nodes represent the server and client within the cluster. 
Throughout our experiments, we assess each plugin in terms of CPU, RAM and throughput across various communication conditions, i.e., i) “Idle”, ii) “Pod-to-Pod”, and iii) “Pod-to-Service”, evaluating both TCP and UDP transport layer protocols. 
In states (ii) and (iii), 
the client Pod establishes a direct connection with the server Pod using its IP address and ClusterIP service, respectively. 

Our experimental runs are categorized into two distinct scenarios, i.e., for intra-host and inter-host communications. In particular, we explore these scenarios in $2$ testbeds, referred to as ATH and UOM, located at the ATHENA Research Center and the University of Macedonia, respectively. 



The ATH testbed includes a single Dell PowerEdge T640 physical machine with an Intel(R) Xeon(R) Silver 4210R at 2.40GHz, $16$ cores CPU, $64$GB RAM and $1$TB SSD, hosting all cluster VMs and demonstrating the intra-host communication scenario. 
On the contrary, the UOM testbed consists of two Dell PowerEdge R630 physical servers each one comprising of $2$ Intel(R) Xeon(R) CPU E5-2620 v4 at 2.10GHz $16$ cores CPU, $68$ \& $34$GB RAM and $1.5$TB HDD. 
The first machine of the UOM testbed hosts the K8s master and one worker node and the second the other worker, i.e., realizing the inter-host communication scenario. 
Both testbeds use the XCP-ng virtualization platform (https://xcp-ng.org). Finally, for each experiment we conduct $10$ iterations of the \textit{knb} tool.

\section{Results}
In this section, we evaluate the quantitative performance of all considered CNI plugins and K8s distributions based on the methodology discussed in the previous section. The results of the two scenarios follow. 

\subsection{Scenario 1 - Intra-host communication}

\begin{figure*}[htb]
\centering
    \begin{subfigure}[b]{\textwidth}
    \includegraphics[height = 2.7cm,width=0.33\textwidth]{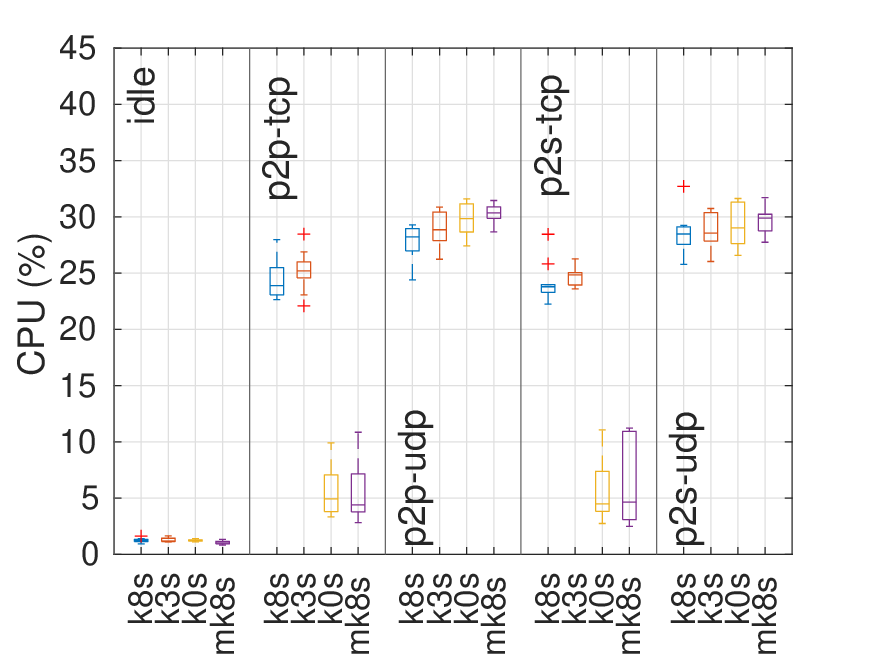}  
    \includegraphics[height = 2.7cm,width=0.33\textwidth]{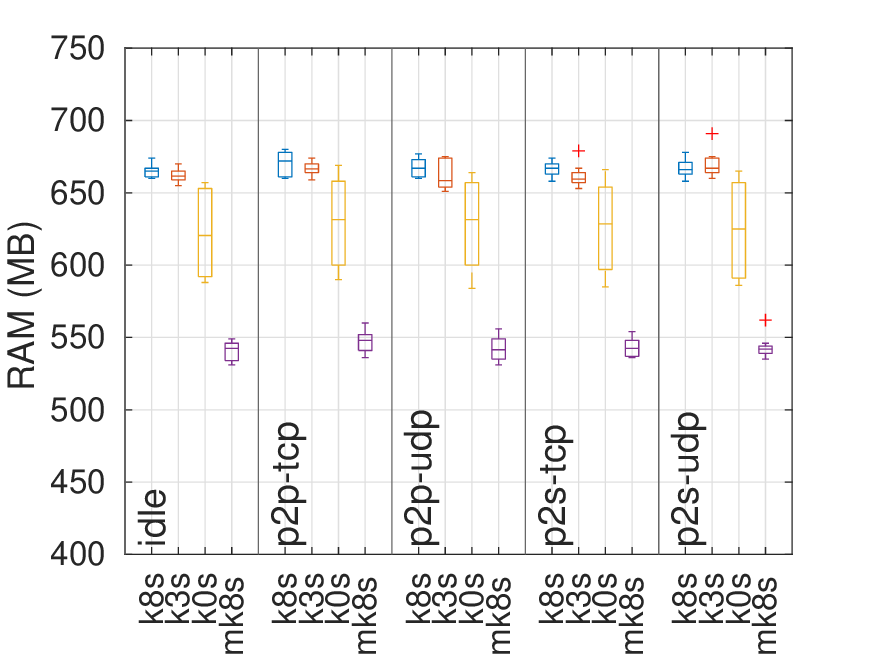}   
    \includegraphics[height = 2.7cm,width=0.33\textwidth]{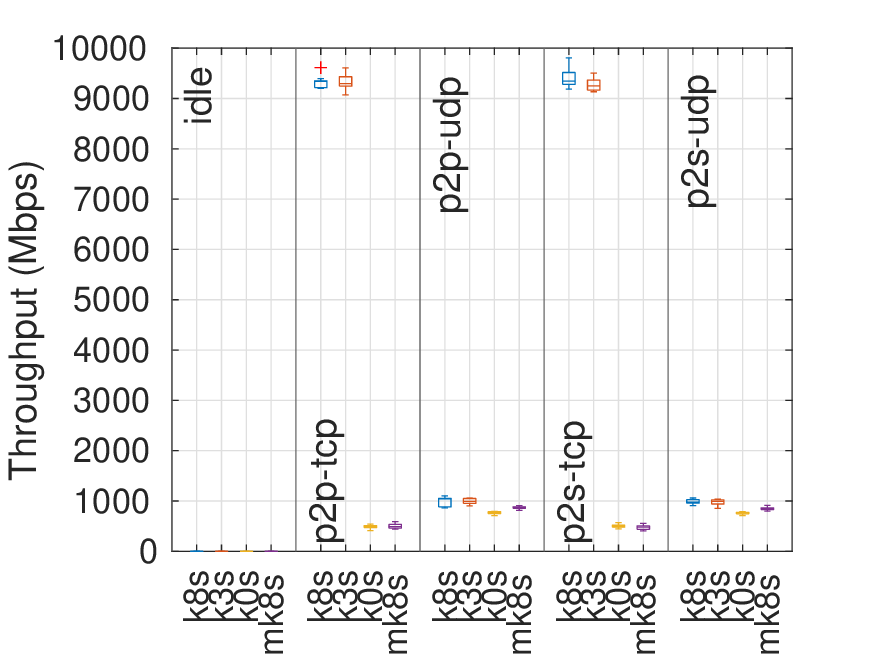}
    \caption{Calico plugin.}
    \end{subfigure}
    
\centering
    \begin{subfigure}[b]{\textwidth}
    \includegraphics[height = 2.7cm,width=0.33\textwidth]{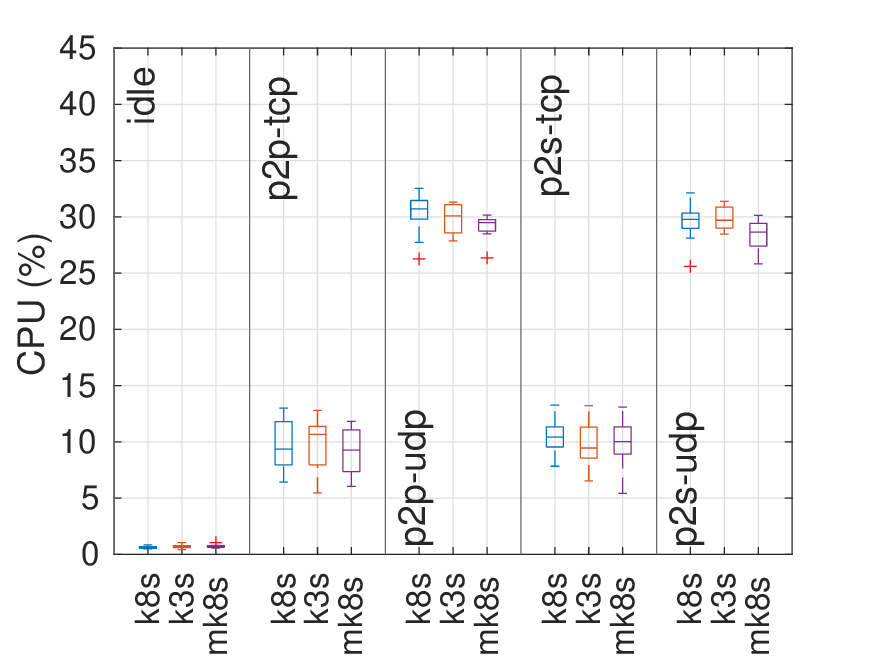}  
    \includegraphics[height = 2.7cm,width=0.33\textwidth]{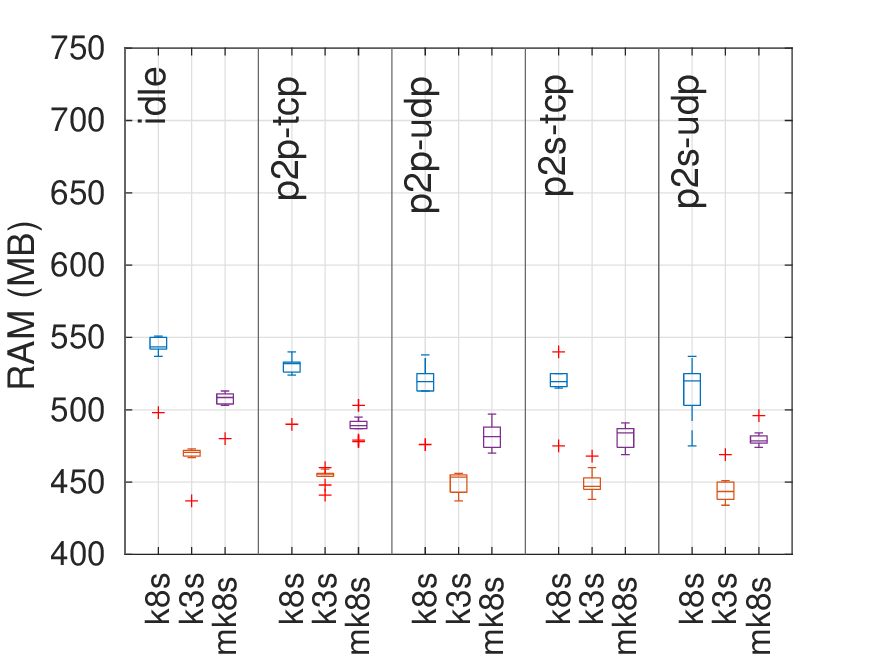}   
    \includegraphics[height = 2.7cm,width=0.33\textwidth]{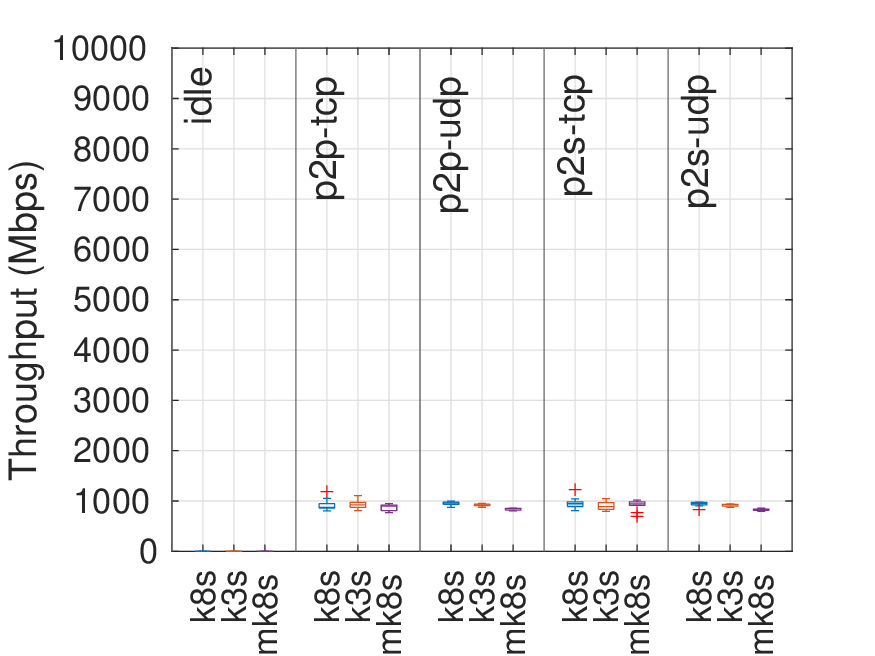}
    \caption{Flannel plugin.}
    \end{subfigure}
    
\centering
  \begin{subfigure}[b]{\textwidth} 
    \includegraphics[height = 2.7cm,width=0.33\textwidth]{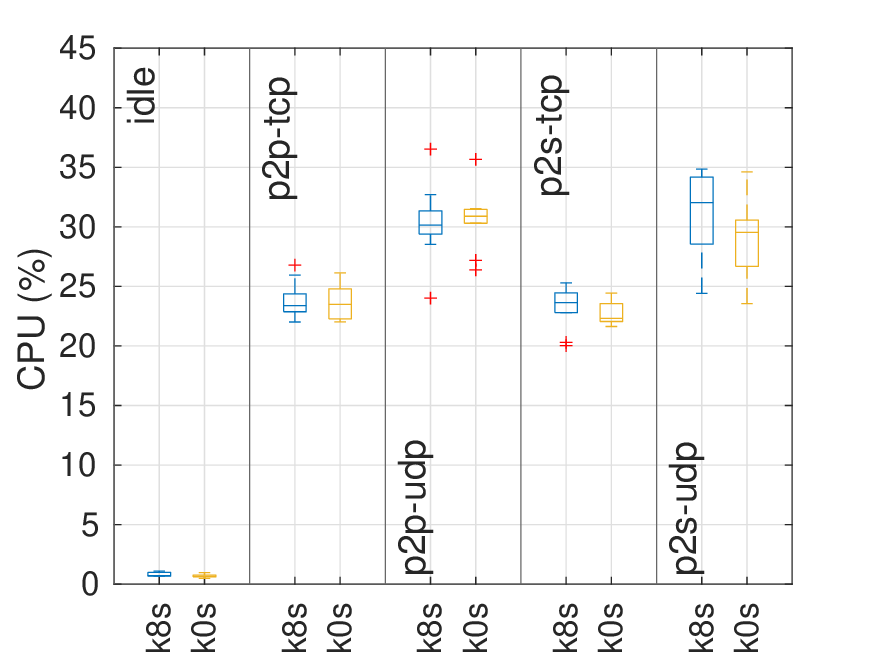}
    \includegraphics[height = 2.7cm,width=0.33\textwidth]{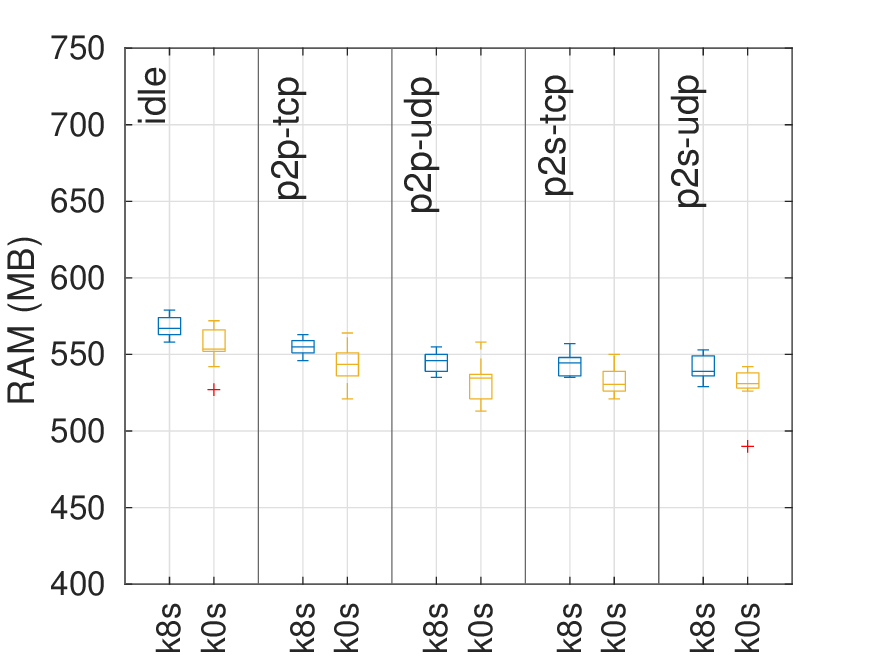}
    \includegraphics[height = 2.7cm,width=0.33\textwidth]{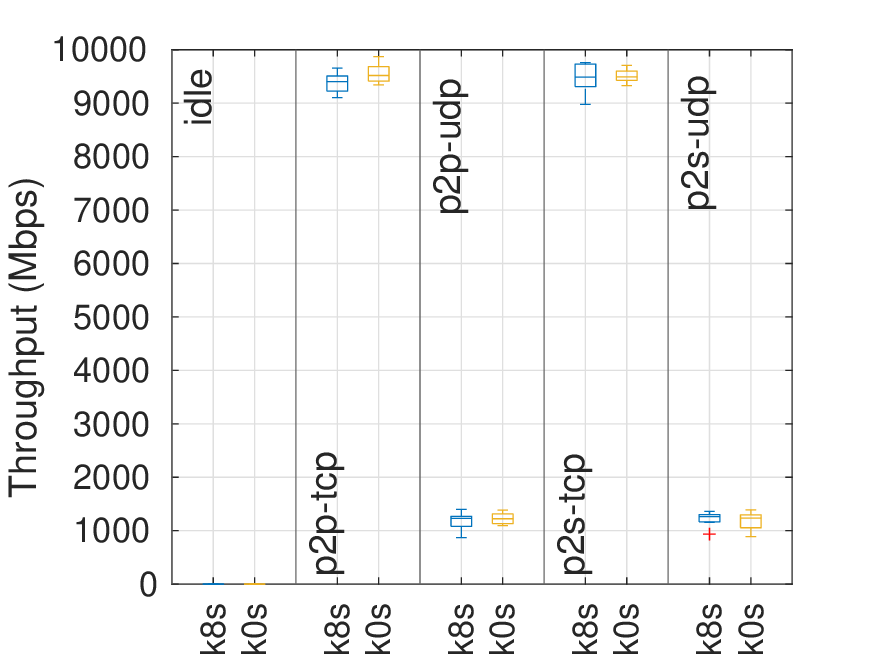}
    \caption{Kube-router plugin.}
  \end{subfigure}
 
\centering
    \begin{subfigure}[b]{\textwidth}       \includegraphics[height = 2.7cm,width=0.33\textwidth]{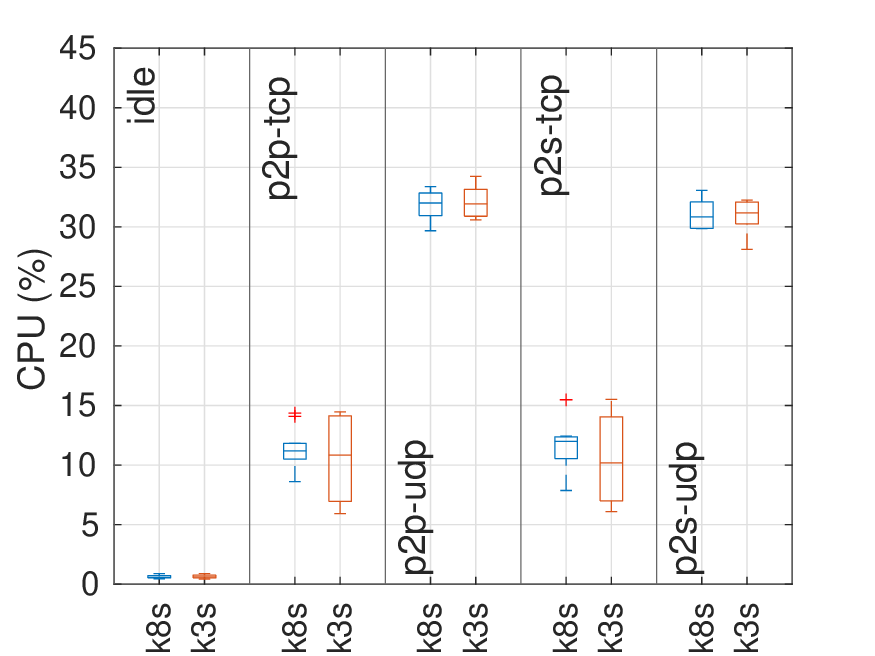}
    \includegraphics[height = 2.7cm,width=0.33\textwidth]{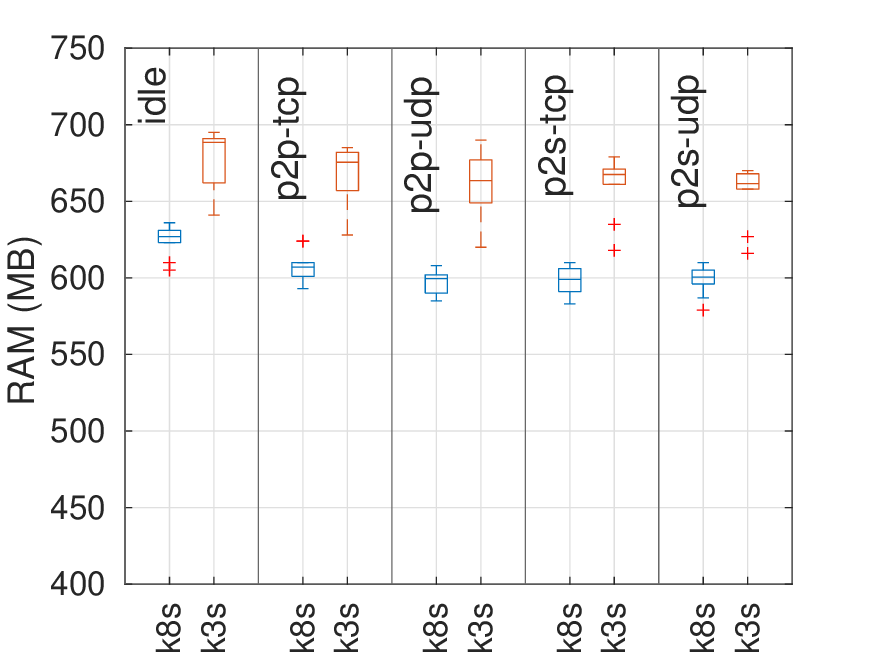}
    \includegraphics[height = 2.7cm,width=0.33\textwidth]{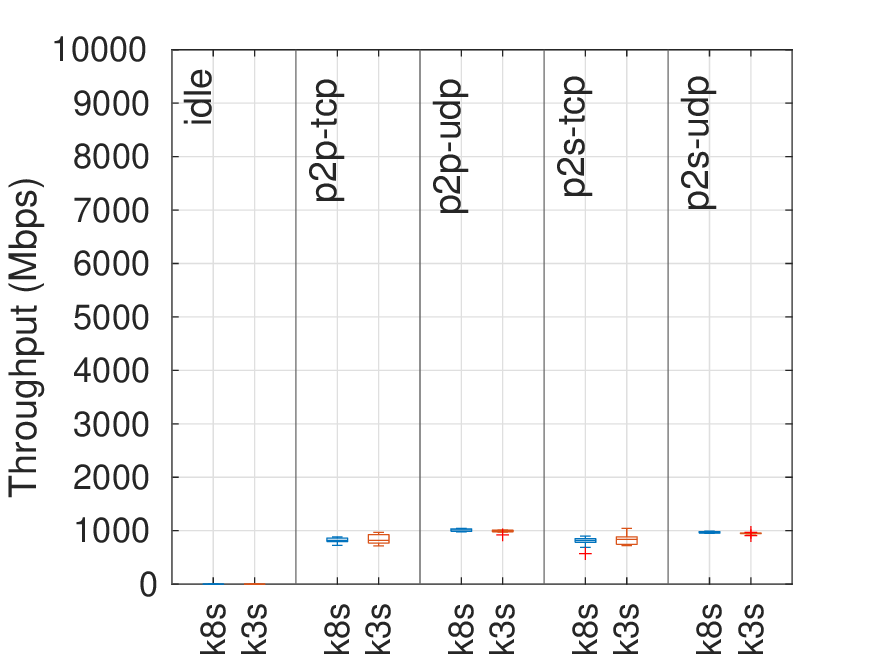}
    \caption{Cilium plugin.}
    \end{subfigure}
    
\centering
    \begin{subfigure}[b]{\textwidth}       \includegraphics[height = 2.7cm,width=0.33\textwidth]{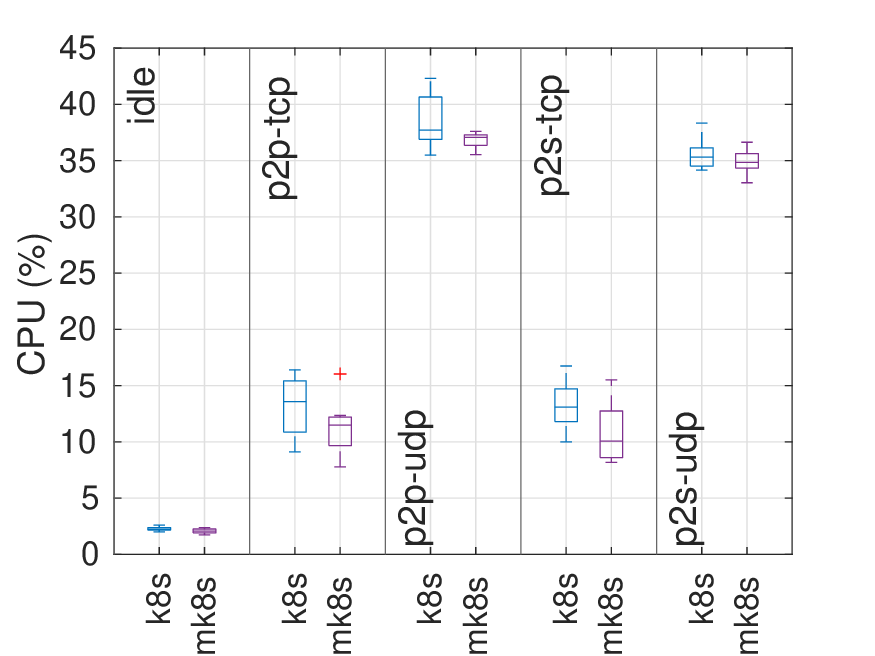}
    \includegraphics[height = 2.7cm,width=0.33\textwidth]{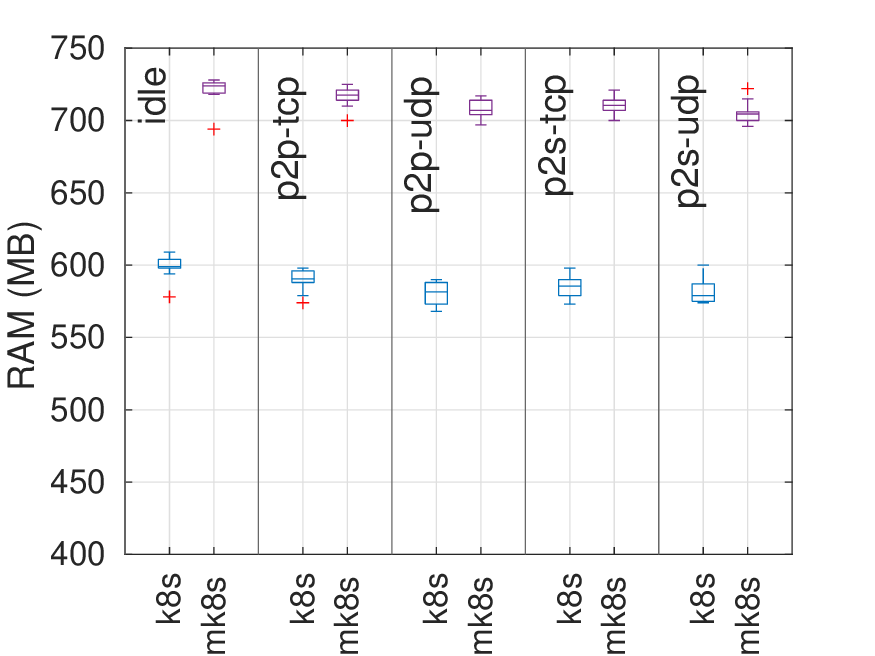}
    \includegraphics[height = 2.7cm,width=0.33\textwidth]{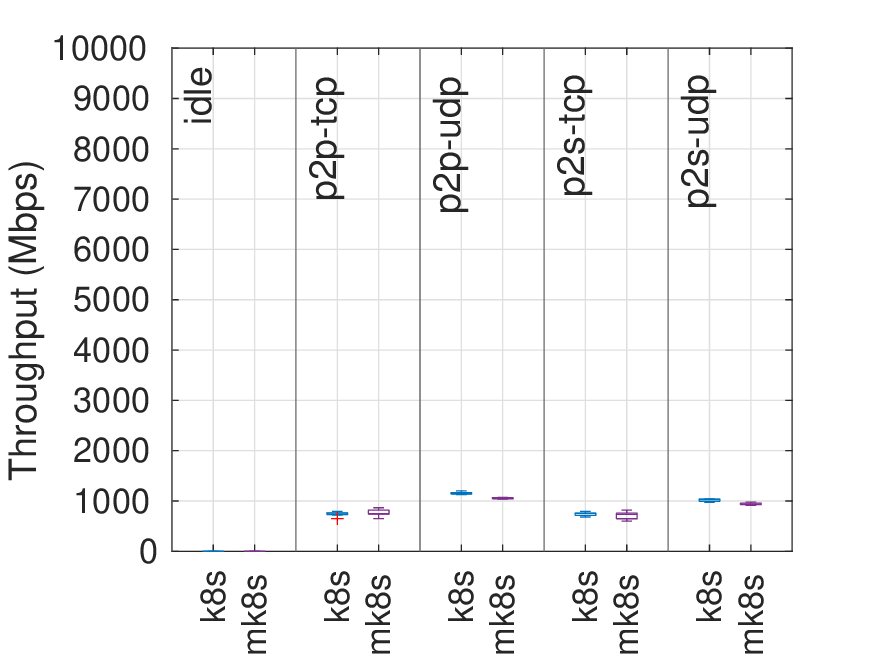}
    \caption{Kube-ovn plugin.}
    \end{subfigure}
    \caption{CPU, RAM usage and throughput per CNI plugin, for different K8s distributions (ATH testbed).}
    \label{fig:ATH_results}
\end{figure*}

Fig. \ref{fig:ATH_results} depicts the resource consumption (CPU, RAM) and the throughput performance of the CNI plugins across the considered K8s distributions, with regards to the intra-host communication scenario (ATH testbed).
Focusing on the CPU usage (leftmost column of Fig. \ref{fig:ATH_results}), the selection of K8s distribution does not significantly impact on the plugins' CPU utilization. An exception is observed for Calico, especially in K0s and Mk8s (in the latter, Calico is the default option), for the TCP communications, where the CPU utilization is reduced by $20\%$ due to a shift from underlay to overlay mechanisms. Furthermore, we notice that UDP is characterized by an increased CPU consumption for all the considered plugins, compared to TCP communication. Focusing on TCP communication, Flannel and Cilium are less CPU intensive, with an average utilization of $10\%$, unlike Calico and Kube-router which show an average CPU consumption of $25\%$. In particular, Flannel offers limited features, while Cilium leverages eBPF hooks to shift processing tasks and provide programmability at the kernel level. 
Concerning UDP communications, besides Kube-ovn that results in $37\%$ CPU utilization, the rest of the deployed plugins provide a CPU consumption of $30\%$. These outcomes are in line with the recent literature, e.g., \cite{Qi} and \cite{Ducastel3}.

On the other hand, the RAM consumption of the employed plugins heavily depends on the K8s distribution in which they are deployed, as shown in Fig. \ref{fig:ATH_results} (middle column). In particular, Flannel efficiently reduces its RAM utilization for both K3s and Mk8s, resulting in the lowest RAM consumption (for the K3s) among the employed plugins. Similarly, Kube-router, the default plugin in the K0s distribution, exhibits the minimum RAM utilization (close to Flannel). However, Kube-ovn and Cilium provide increased RAM consumption when deployed over lightweight K8s distributions, i.e., Mk8s and K3s, respectively. The illustrated results indicate that the selection of a lightweight K8s distribution does not always ensure a lower resource consumption for the plugins. For example, in case of Kube-ovn one has to consider the impact of an SDN approach over resource constrained environments or the increased memory utilization of Cilium due to the footprint of both daemonset and binary files \cite{Qi}.
Moreover, the reduced RAM usage of Calico for the K0s and Mk8s distributions aligns with the results of CPU utilization. We note that the selection of transport layer protocols, i.e., UDP or TCP, does not affect the RAM consumption of the considered plugins.

Finally, the rightmost column of Fig. \ref{fig:ATH_results} illustrates the throughput performance. In the TCP-related conditions, the pure IP-based leading plugins, i.e., Calico and Kube-router, achieve throughput performances that exceed $9000$ Mbps, whereas the remaining plugins consistently fall at $1000$ Mbps. An exception is identified for Calico (in K0s and MicroK8s), where the reduction in resource consumption comes at the cost of decreased throughput. On the other hand, UDP-based communications exhibit lower throughput performance near $1000$ Mbps for all plugins across the different distributions. 

%

\begin{figure*}[htb]
\centering
    \begin{subfigure}[b]{\textwidth}
    \includegraphics[height = 2.7cm,width=0.33\textwidth]{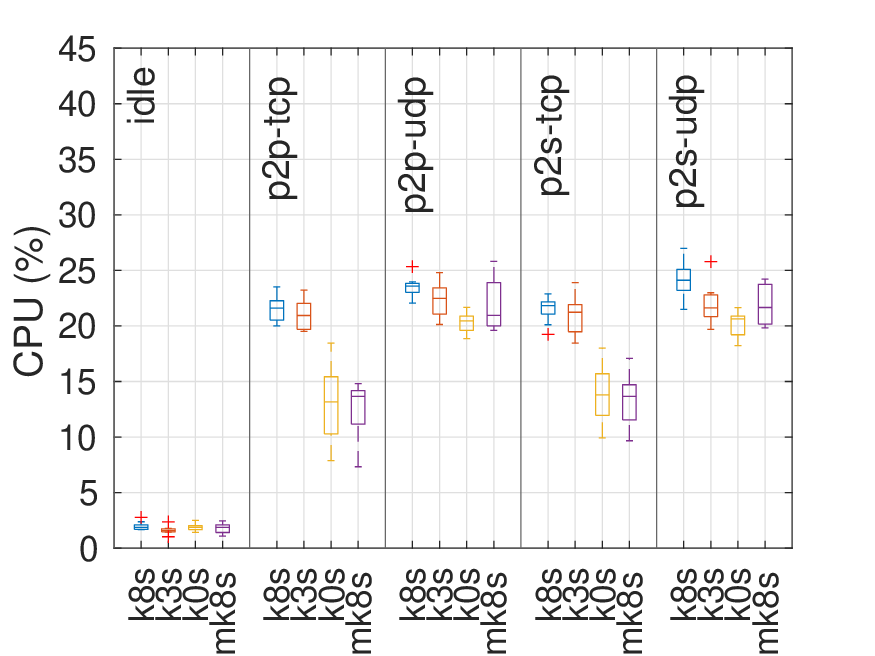}  
    \includegraphics[height = 2.7cm,width=0.33\textwidth]{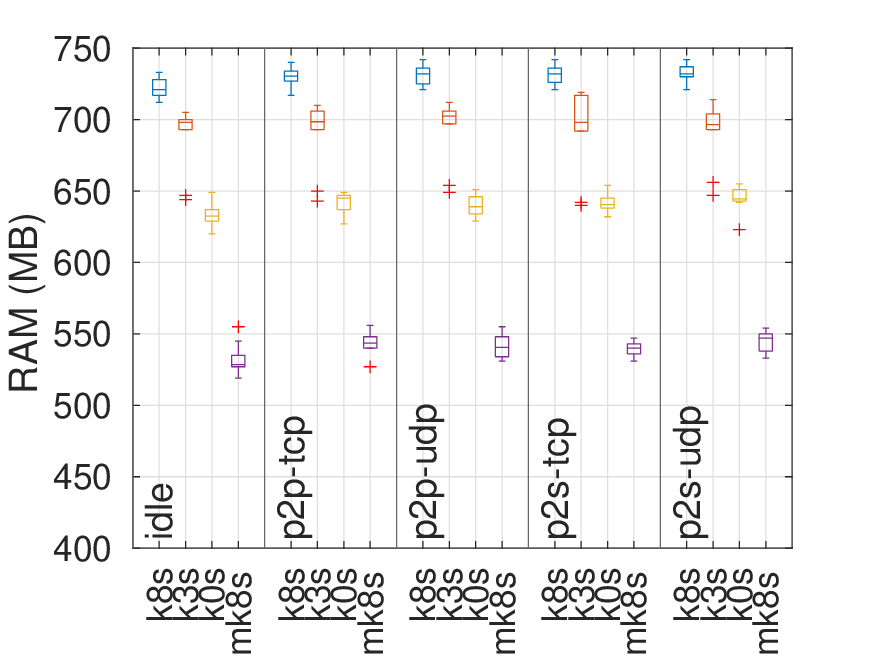}   
    \includegraphics[height = 2.7cm,width=0.33\textwidth]{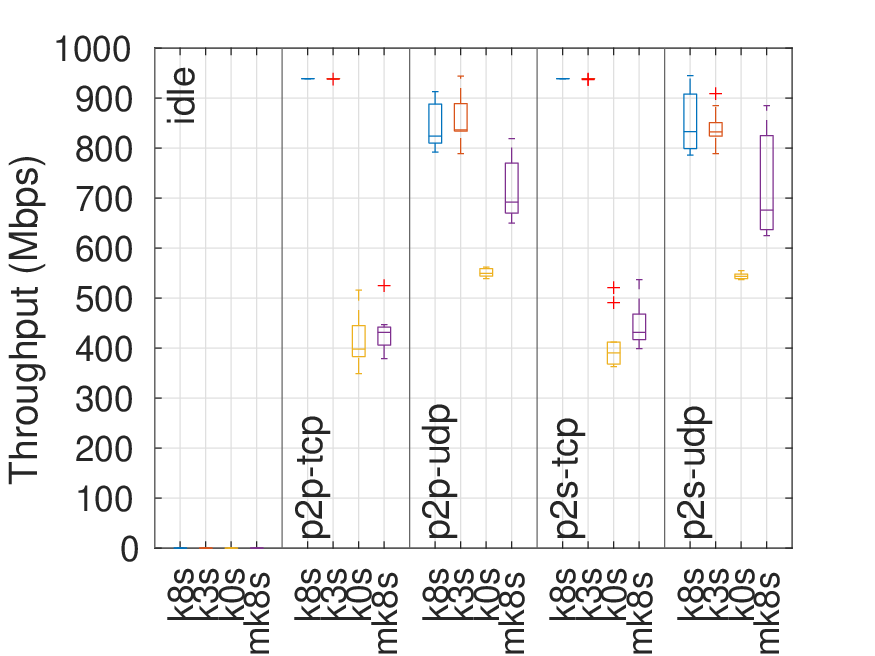}
    \caption{Calico plugin.}
    \end{subfigure}
    
\centering
    \begin{subfigure}[b]{\textwidth}
    \includegraphics[height = 2.7cm,width=0.33\textwidth]{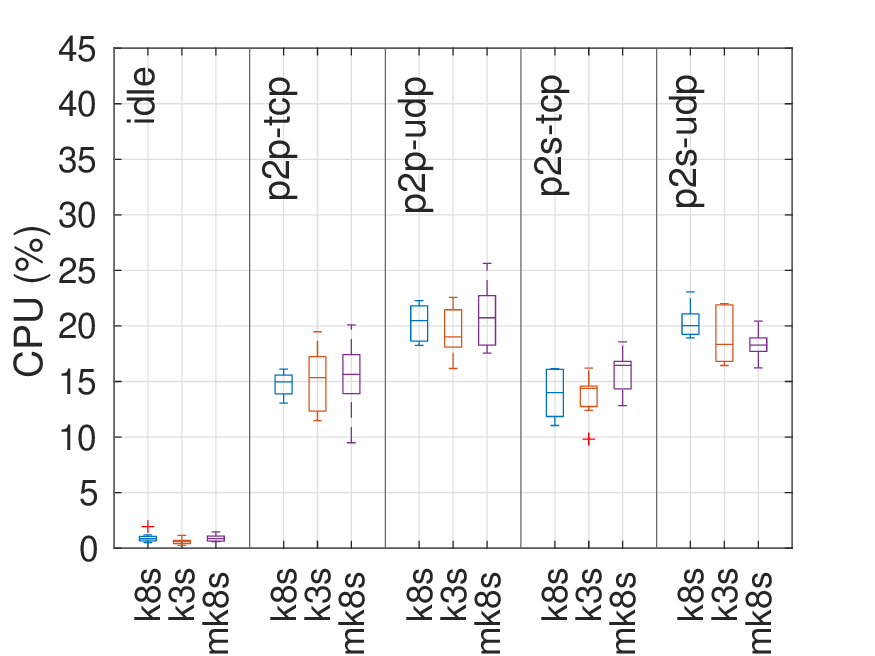}  
    \includegraphics[height = 2.7cm,width=0.33\textwidth]{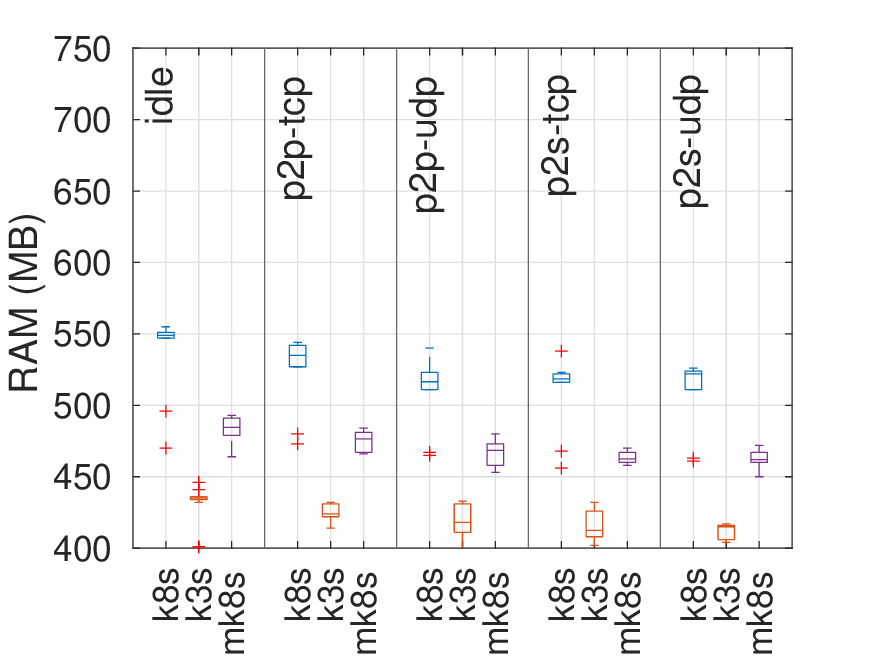}   
    \includegraphics[height = 2.7cm,width=0.33\textwidth]{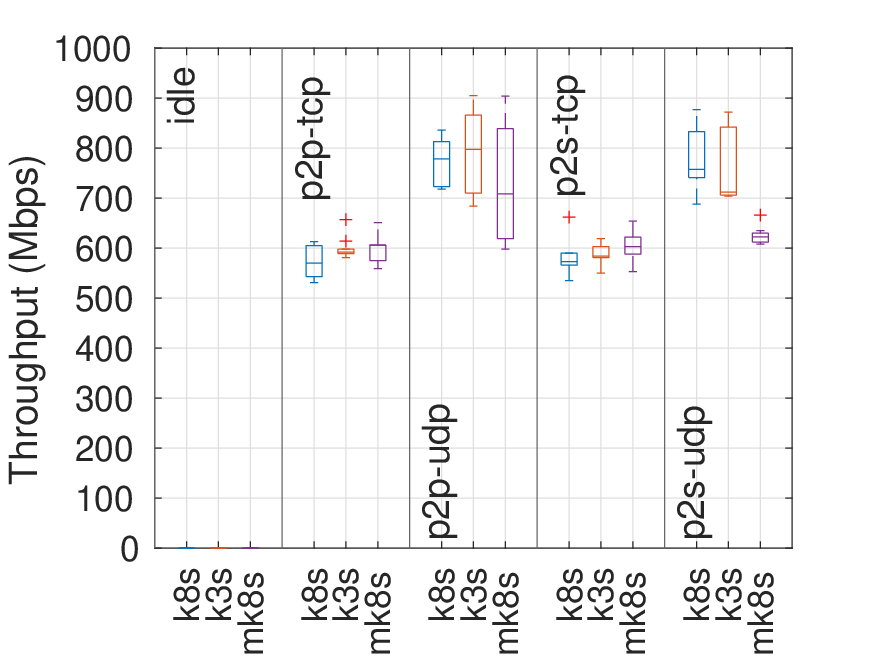}
    \caption{Flannel plugin.}
    \end{subfigure}
    
\centering
  \begin{subfigure}[b]{\textwidth} 
    \includegraphics[height = 2.7cm,width=0.33\textwidth]{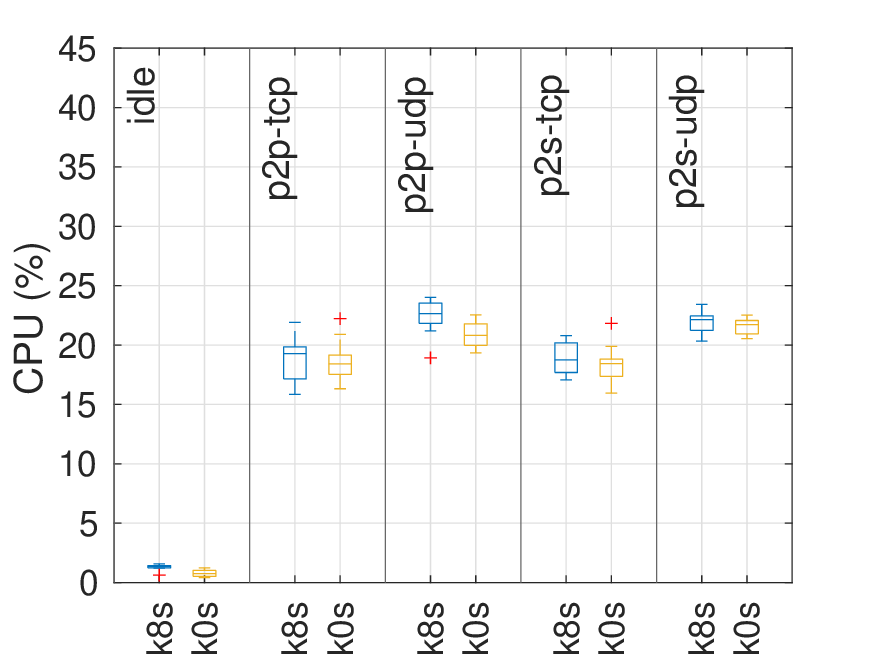}
    \includegraphics[height = 2.7cm,width=0.33\textwidth]{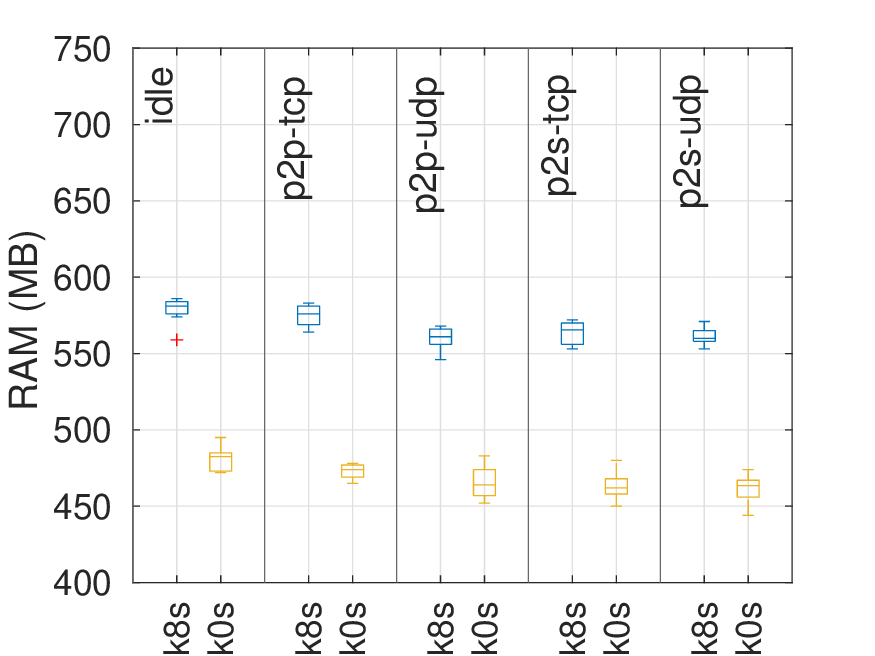}
    \includegraphics[height = 2.7cm,width=0.33\textwidth]{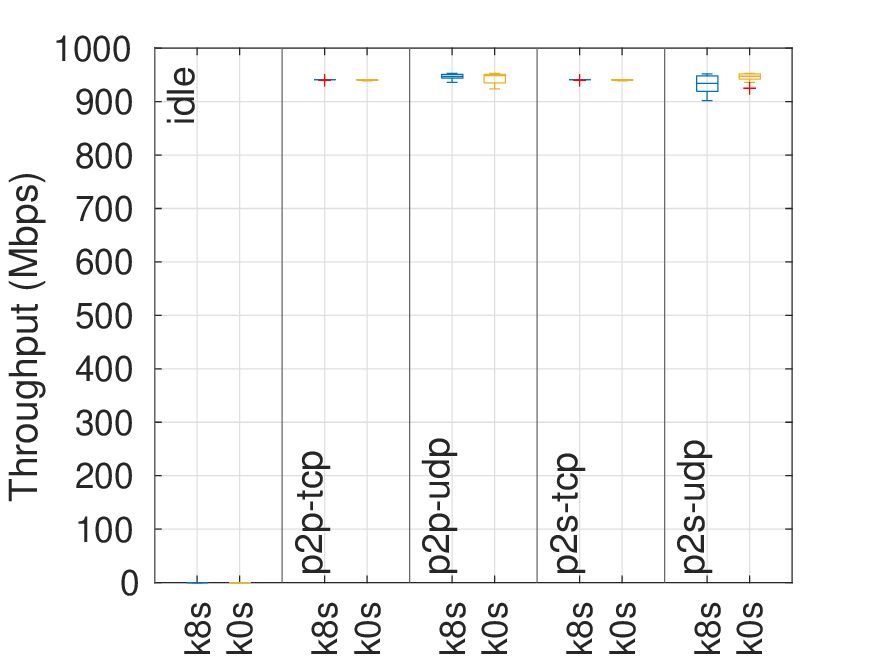}
    \caption{Kube-router plugin.}
  \end{subfigure}
 
\centering
    \begin{subfigure}[b]{\textwidth}       \includegraphics[height = 2.7cm,width=0.33\textwidth]{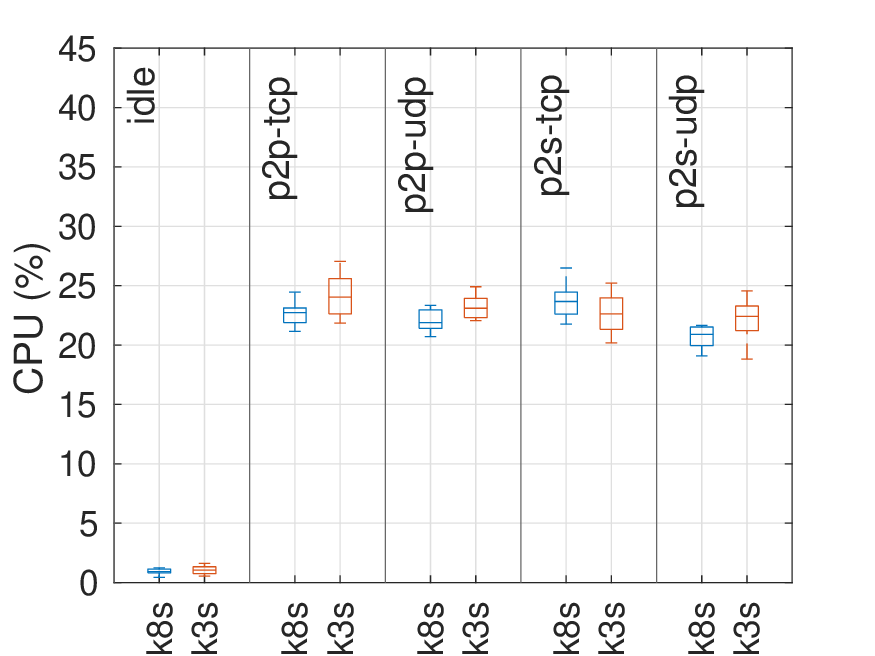}
    \includegraphics[height = 2.7cm,width=0.33\textwidth]{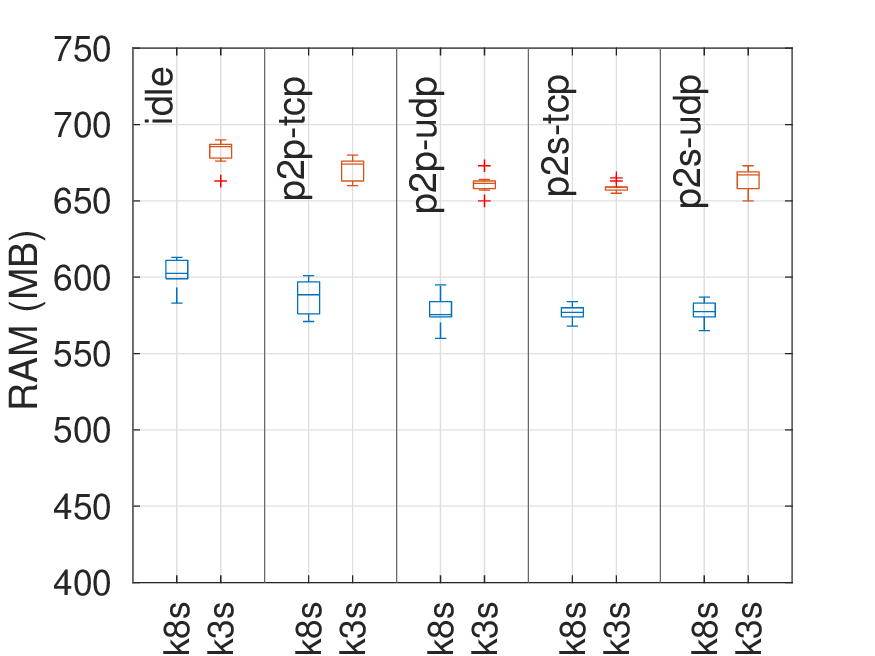}
    \includegraphics[height = 2.7cm,width=0.33\textwidth]{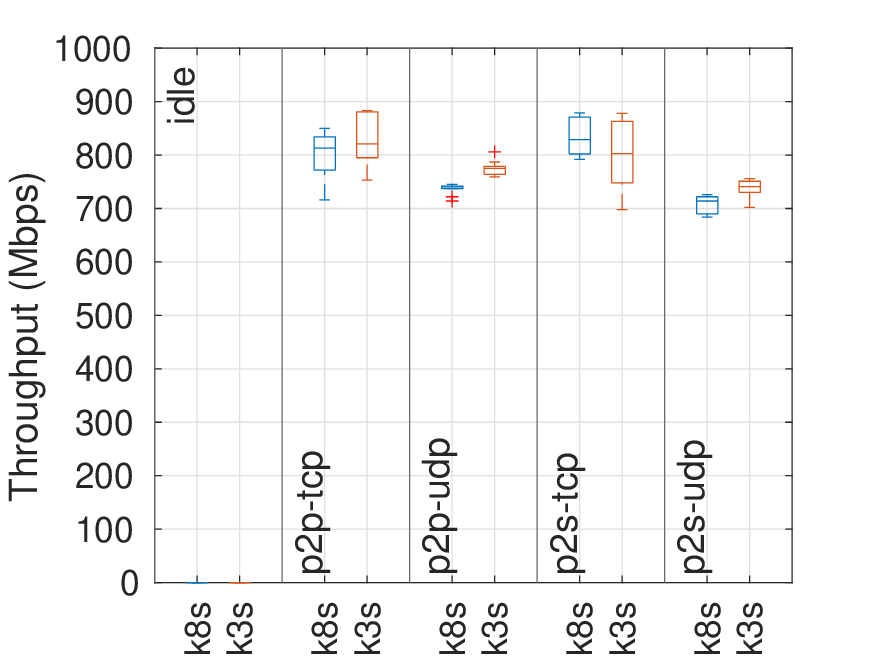}
    \caption{Cilium plugin.}
    \end{subfigure}
    
\centering
    \begin{subfigure}[b]{\textwidth}       \includegraphics[height = 2.7cm,width=0.33\textwidth]{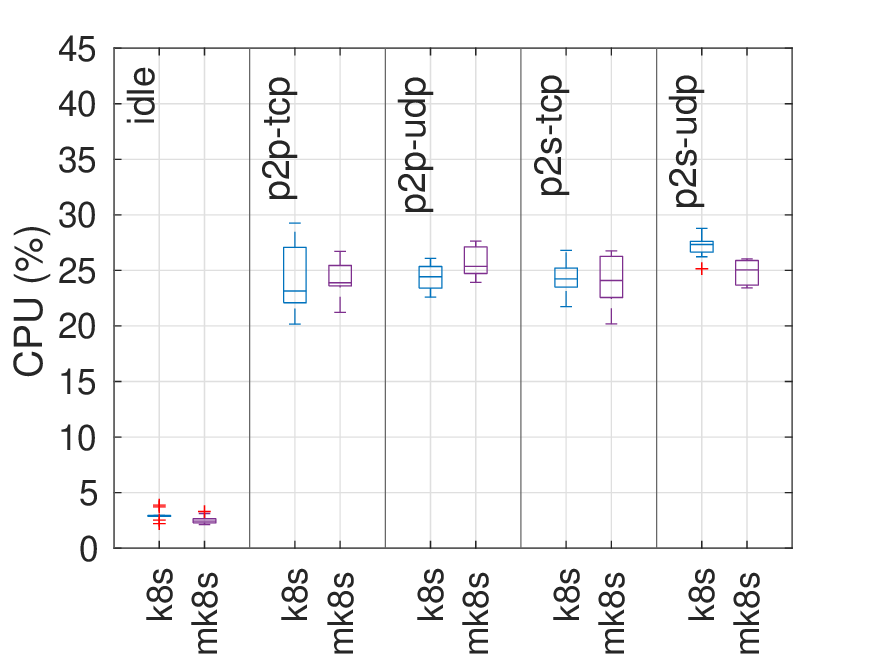}
    \includegraphics[height = 2.7cm,width=0.33\textwidth]{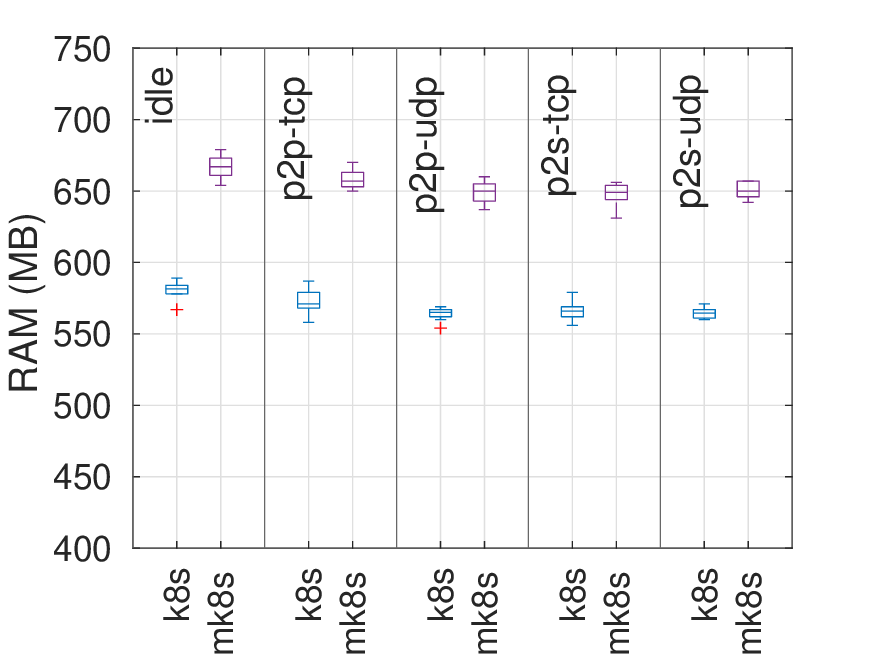}
    \includegraphics[height = 2.7cm,width=0.33\textwidth]{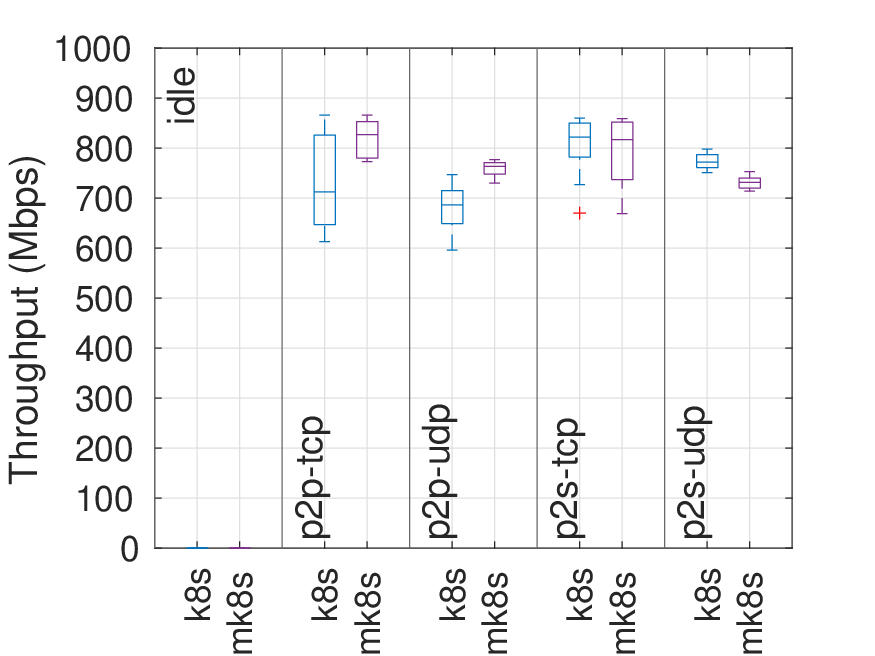}
    \caption{Kube-ovn plugin.}
    \end{subfigure}
    \caption{CPU, RAM usage and throughput per CNI plugin, for different K8s distributions (UOM testbed).}
    \label{fig:UOM_results}
\end{figure*}

\subsection{Scenario 2 - Inter-host communication}
In Fig. \ref{fig:UOM_results}, we investigate the performance of the deployed plugins in the inter-host communication scenario (UOM testbed). 
Regarding the CPU usage across K8s distributions (Fig. \ref{fig:UOM_results}, leftmost column), the results are aligned with those observed in the intra-host scenario, since the K8s distribution does not seem to significantly affect the CPU consumption of the plugins (again with the exception of Calico, for K0s and Mk8s). On the other hand, the inter-host communication results in lower performance variations between UDP and TCP communication. In general, the CPU is increased for the TCP and decreased for the UDP communication, Kube-router and Calico being excepted. 
Flannel outperforms the rest of the plugins in terms of CPU consumption, while Kube-ovn is the most CPU-intensive, due to its inherent ``rich" and programmable features.

In terms of RAM usage (middle column of Fig. \ref{fig:UOM_results}), the results reveal similar patterns to the intra-host scenario, for all plugins. For instance, compared to vanilla K8s distribution, Flannel reduces its RAM footprint by $28\%$ in K3s and $15\%$ in Mk8s. Kube-router by 21\% over K0s (in which it is the default plugin), while Calico by $6\%$, $16\%$ and $37\%$, regarding K3s, K0s and Mk8s. On the contrary, Kube-ovn and Cilium result in $16\%$ and $14\%$ increased RAM consumption, for Mk8s and K3s distribution, respectively.

Finally, the throughput performance of the plugins is shown in Fig. \ref{fig:UOM_results} (rightmost column). As expected, the inter-host scenario presents decreased throughput performance, e.g., no plugin exceeds the $1000$ Mbps threshold. Moreover, plugins based on underlay IP routing, i.e., Calico and Kube-router, surpass the performance of VxLAN overlay-based ones, i.e., Flannel and Cilium. In this context, Kube-router exhibits higher and more consistent throughput performance, for UDP/TCP protocols. On the other hand, the lightweight nature of Flannel impacts its throughput performance, especially regarding the TCP protocol. Also, we notice a decrease in the performance of Calico for K0s and MicorK8s, as in the intra-host scenario. 
Finally, the increased resource consumption of Kube-ovn and Cilium, when deployed over lightweight K8s distributions, seems to offer no performance gains in terms of throughput.  

\section{Discussion}
Based on the provided qualitative and quantitative analysis, we summarize our key insights as follows: 

(i) The CPU consumption remains relatively stable across the plugins, regardless of their deployment 
over vanilla K8s or a lightweight distribution. An exception is identified for Calico in K0s and Mk8s due to its shift from underlay to overlay mechanisms.
Overall, the results underscore the necessity for further fine-tuning and optimization of plugins in Edge environments, wherein reducing the CPU consumption is normally desired due to the associated resource constraints.

(ii) The RAM usage is generally reduced on lightweight K8s distributions, except for Cilium and Kube-ovn plugins. In this regard, Flannel 
provides the lowest resource footprint, particularly excelling across lightweight K8s. However, it lacks in terms of throughput and advanced features (e.g., security, network policies) which impacts its applicability in actual production environments. Conversely, Kube-ovn and Cilium, which support advanced programmable features, exhibit a performance decline in terms of RAM consumption and throughput in lightweight distributions. Such a trade-off should be considered upon plugin selection to align with the specific requirements of the implementation scenario.

(iii) The reduction of resource footprint over Edge distributions does not necessarily impact the throughput of the considered plugins, with the exception of Calico. 

(iv) According to our evaluation results, Calico, Flannel and Kube-router adjust their resources across lightweight distributions, with the former being used as the default plugin in K0s. Calico, specifically, offers advanced functionalities and manages to significantly reduce its resource utilization, however at the cost of decreased throughput performance. 
 

\section{Conclusion and Future Work}
We presented 
a qualitative and quantitative performance evaluation of commonly used K8s networking solutions across various K8s distributions, emphasizing on their performance across resource-restricted Edge environments. Our experiments over intra/inter-host communication scenarios revealed that lightweight K8s implementations mainly reduce the RAM footprint of certain plugins (i.e., Calico, Kube-router and Flannel), usually at the cost of throughput, especially for Calico and Flannel. However, the employment of lightweight K8s distributions does not ensure a reduction in the resource consumption of plugins, e.g., Cilium and Kube-ovn, highlighting the need for further optimizations across Edge-related distributions. Our future plans include the investigation of: (i) out-of-the-box plugins, e.g., L2S-M (https://github.com/Networks-it-uc3m/L2S-M), (ii) additional performance metrics, e.g., distributions' life-cycle and scalability aspects, and (iii) the performance of typical applications deployed at the Edge.


\section*{Acknowledgment}
This work is supported by the Horizon Europe CODECO Project under Grant number 101092696.

\bibliographystyle{IEEEtran}
\bibliography{conference_101719.bib}

\end{document}